# ON THE INTERACTION OF THE ULTRA-SHORT LASER PULSES WITH HUMAN BRAIN MATTER

M. Kozlowski[1,*], J. Marciak – Kozlowska[2]

[1] Physics Department, Warsaw University, Warsaw, Poland
[2] Institute of Electron Technology, Warsaw, Poland

**Abstract**

In this paper the modified Schrodinger equation for attosecond laser pulses interaction with "atoms" of human brain . i.e. neurons is developed and solved. Considering that the mass of the human neuron is of the order of Planck mass $=10^{-5}$ g the model equation for Planck masses is applied to the laser pulse neuron interaction

corresponding author: e-mail: miroslawkozlowski@aster.pl

1. Introduction

Over recent years there has been a flourish of research furthering our understanding of brain function, connectivity, hemodynamics, chemistry, mapping etc; this progress has been made hand in hand with the development of a plethora of techniques to probe or observe the brain in action. These techniques range from fMRI, PET and Laser speckle imaging to optical intrinsic signal imaging and near-infrared spectroscopy amongst others Each technique adds its portion of information to the puzzle but also brings with it its own specific technical limitations. The surgical removal of a lesion adjacent to motor, sensory, language or visuospatial areas requires their identification in order that these functionally-important regions of the brain may be preserved. Neurosurgery has gained considerably in precision over recent years thanks to the availability of methods for visualizing the brain and its functioning, such as the use of fMRI] for brain mapping, the integra tion of fMRI into neuronavigation tools for use during surgery together with cortical electrostimulation . FMRI results are complicated by 'brain shift' necessitating the recalculation of navigation data during theatre, and by the large vessel effect extending the signal towards the draining vessels. Electro-cortical stimulation is the surgical gold standard and directly reveals interference with brain function, but it is a laborious and time consuming technique yielding occasional false positives and exposing the patient to the risk of seizure. Thus there remains need for a rapid, high resolution, non- invasive method applicable in theatre allowing the visualization of areas of cortical activation across the exposed brain, i.e. a wide held of view, as well as enabling many different and complex functional maps to be generated intraoperatively.

Femtosecond lasers have been well established as a tool for subsurface machining of transparent materials [1] and ultra-precise machining of solid state materials [2]. Recently, femtosecond ablation has been adapted as a biological tool to manipulate structures and chemistry on a microscopic scale. Femtosecond laser ablation can produce ionization that is highly spatially confined in three dimensions. This can be used to disrupt organelles inside individual cells with subcellular precision, multiple cells, or parts of cells in whole animals. This allows unprecedented study of biological dynamics and reactions. These capabilities allow the function of a particular structure to be studied in an *in vivo* context, and can also be used to reproduce disease states of relevance to human health in animal models.
At ordinary intensities of light, photons that have insufficient energy to ionize a material pass through the material without interaction; for example, visible light passes through transparent materials such as water and glass. However, at light intensities around $10^{13}$ W/cm$^2$ that are achievable by femtosecond lasers, the density of photons is sufficient that the chance of multiple photons interacting with the same molecule nearly simultaneously becomes high. The cooperative action of multiple photons can result in ionization of an ordinarily transparent material via several conceptually different processes . In multiphoton ionization, an unexcited electron simultaneously absorbs the energy of several photons, so that the combined energy of these photons is sufficient to boost the electron to an excited state or may free the electron to form an electron-ion plasma. Because overcoming the energy gaps in transparent materials require multiple photons, the probability of ionization depends on intensity to a power equal to the number of photons. This process requires that the photon density is high enough that the probability of absorbing multiple photons at one time is non-negligible. Another mechanism for cooperative action of photons is called avalanche ionization. In this process a previously ionized electron can absorb energy from the laser photons. Collisions between this electron and an un-ionized electron can result in the ionization of the second electron. This process results in an exponentially increasing number of ionized electrons. Avalanche ionization also requires high laser intensities because it requires that an ionized electron absorb the energy of multiple photons so that its excess energy exceeds the band-gap energy before colliding with another electron.

Ionization leads to an electron-ion plasma in which the energy from the laser pulse is stored in the separation of the positively charged ions and negative electrons and their kinetic energy. After about 10 ps to 1 ns, the electrons and ions in the plasma recombine. The energy from the recombination contributes to highly non-equilibrium conditions, which can overcome the tensile force of the material around the focus . With sufficient energy, this leads to a microexplosion and shockwave [3,4] The recombined material, including molecular hydrogen and oxygen [5], remains as a hot gas bubble. This cavitation bubble expands and contracts with complex dynamics as it equilibrates to the surrounding conditions. All the steps in the process of femtosecond laser ablation can produce effects in biological materials. For example, the material ionized at the laser focus may be altered substantially from its original chemical composition. Ionization can also lead to excited molecules that participate in chemical reactions. Both the formation of reactive oxygen species from the ionization of water and the direct breaking of chemical bonds in cellular structures can lead to biological effects [6] The microexplosion, shockwave, and cavitation bubble can add mechanical disruption to chemical effects [7,8]

The use of femtosecond lasers as a precise, subsurface scalpel has also been extended to the worm *C. elegans*, in which changes in behavior in whole active animals have been observed after laser ablation. Yanik. [7] used pulses to cut single axons in neurons that are responsible for coordinating muscles that cause the worm to move backwards after a touch on the nose. Femtosecond lasers are ideal tools for this research because the subsurface ablation capability enabled work in whole animals, allowing the observation of behavioral changes. In addition, the microscopically-sized affected volume allowed this research team to cut a single axon at a time. In animals in which multiple axons were cut lost the ability to go backwards after being touched on the nose. Remarkably, the laser ablation was mild enough that over half the cut neurons showed signs of regrowth within 24 hours. The majority of animals with cut neurons recovered some ability to move backwards after 24 hours, indicating that at least some of the axons had recovered functionality. Other techniques for eliminating neurons in *C. elegans* can disrupt whole cells. However, after elimination of the whole neuron, rather than just the axon, the worm does not recover the ability to move backwards even after 48 hours. The femtosecond laser ablation helped elucidate difference between losing a portion of a cell, rather than the entire cell.

Femtosecond laser ablation has also been used to elucidate the role of specific neurons involved in sensory systems in *C. elegans* .. It had been previously known that a group of neurons were involved in the regulating behaviors related to temperature. *C. elegans* is known to be sensitive both to absolute temperature and also to changes in temperature. In the work from Chung [9].individual dendrites in the neural network responsible for temperature sensing were cut by femtosecond laser, and the reaction of the worms to different temperatures and temperature gradients was measured. From these dissections, the authors were able to determine that a particular neuron is involved in generating a response to absolute temperature, but is not involved regulating behavior in response to relative changes in temperature. Both of these experiments are remarkable in that they relate the physical "wiring" of a neural network all the way to behavior.

In this paper we will investigate the new Schrodinger equation for the analysis of the attosecond laser pulse interaction with human brain, i.e. with the net of the neurons.

Schrodinger equation describes the extraordinary behavior of the matter and energy which comprise our Universe at a fundamental level. At the root of Schrodinger equation is the wave/particle duality of atoms, molecules and their constituting particles. A quantum system such as an atom or sub-atomic particle which remains isolated from its environment behaves as a wave of possibilities and exists in a coherent complex number valued "superposition" of many possible states. The behavior of such wave-like quantum level objects can be satisfactorily described in terms of state vector which evolves deterministically according to the Schrödinger equation (unitary evolution) denoted by U.

Somehow quantum microlevel superposition leads to unsuperposed stable structures in our macro-world. In a transition known as a wave function collapse or reduction (R), the quantum wave to alternative possibilities reduces to a simple macroscope reality, an "eigenstate" of some appropriate operator [10].

In seminal paper [11] Sultan Tarlaci described the strong connection and mutual interaction of quantum theory and cognitive science. The master equation of quantum theory is the Schrödinger equation. In this paper we intend to biologize Schrödinger equation in order to find out the strongest relation between neuroscience and quantum physics.

In our paper [13] we developed new form of Schrödinger equation. Modified Schrödinger Equation (MSE) with new term which describes the "memory" of the quantum state. In this paper we will show that memory of all quantum state is influenced by gravity.

As was shown in paper [10] the R states of quantum physics are originated at Planck level. In this paper we underline the fact that the Planck mass is of the order of the mass of the human neuron, $M_P = m_{HN} = 10^{-5}$ g. It is interesting to observe that Planck mass is the mass at which the modified Schrödinger equation changes the structure from parabolic for mass $m < M_P = m_{HN}$ to hyperbolic for mass $m > M_P = m_{HN}$.

2. Modified Schrödinger Equation

When M. Planck made the first quantum discovery he noted an interesting fact [12]. The speed of light, Newton's gravity constant and Planck's constant clearly reflect fundamental properties of the world. From them it is possible to derive the characteristic mass $M_P$, length $L_P$ and time $T_P$ with approximate values

$L_P = 10^{-35}$ m
$T_P = 10^{-43}$ s
$M_P = 10^{-5}$ g.

The enigma of the "classical "value of the Planck mass is not solved by existing astro-and physical theories. Considering that typical masses of the quantum particles are of the order of $10^{-27}$g – the mass of the order of $10^{-5}$ g is quite astonishing for the fundamental particle of the Planck Epoch.

In this paper considering the structure of the human brain we note that the mass of the neuron- the " atom" of human brain is the order of the $10^{-5}$ g. In the papers [ 3-15 ] we developed the model for thermal phenomena in Planck gas, i.e. the gas of particles with the mass=$10^{-5}$ g

Considering the results of the paper [ 12 ] we can applied the model equations to the study of the transport processes in human brain. The Planck mass is equal $M_p = (hc/G)^{1/2}$ and depends on the gravity constant G

First of all let us consider the question: how gravity can modify the quantum mechanics, i.e. the nonrelativistic Schrödinger equation (SE). We argue that SE with relaxation term describes properly the quantum behavior of particle with mass $m_i < M_P$ and contains the part which can be interpreted as the pilot wave equation. For $m_i \quad M_P$ the solution of the SE represent the strings with mass $M_P$.

The thermal history of the system (heated gas container, semiconductor or Universe) can be described by the generalized Fourier equation [12]

$$q(t) = -\int_{-\infty}^{t} \underbrace{K(t-t')}_{\text{thermal history}} \underbrace{\nabla T(t')dt'}_{\text{diffusion}}. \qquad (1)$$

In Eq. (1) $q(t)$ is the density of the energy flux, $T$ is the temperature of the system and $K(t - t')$ is the thermal memory of the system

$$K(t-t') = \frac{K}{\tau}\exp\left[-\frac{(t-t')}{\tau}\right], \qquad (2)$$

where $K$ is constant, and $\tau$ denotes the relaxation time.
As was shown in [2]

$$K(t-t') = \begin{cases} K\delta(t-t') & \text{diffusion} \\ K = \text{constant} & \text{wave} \\ \frac{K}{\tau}\exp\left[-\frac{(t-t')}{\tau}\right] & \text{damped wave} \\ & \text{or hyperbolic diffusion}. \end{cases}$$

The damped wave or hyperbolic diffusion equation can be written as:

$$\frac{\partial^2 T}{\partial t^2} + \frac{1}{\tau}\frac{\partial T}{\partial t} = \frac{D_T}{\tau}\nabla^2 T. \tag{3}$$

For $\tau \to 0$, Eq. (3) is the Fourier thermal equation

$$\frac{\partial T}{\partial t} = D_T \nabla^2 T \tag{4}$$

and $D_T$ is the thermal diffusion coefficient. The systems with very short relaxation time have very short memory. On the other hand for $\tau \to \infty$ Eq. (3) has the form of the thermal wave (undamped) equation, or *ballistic* thermal equation. In the solid state physics the *ballistic* phonons or electrons are those for which $\tau \to \infty$. The experiments with *ballistic* phonons or electrons demonstrate the existence of the *wave motion* on the lattice scale or on the electron gas scale.

$$\frac{\partial^2 T}{\partial t^2} = \frac{D_T}{\tau}\nabla^2 T. \tag{5}$$

For the systems with very long memory Eq. (3) is time symmetric equation with no arrow of time, for the Eq. (5) does not change the shape when $t \to -t$.

In Eq. (3) we define:

$$v = \left(\frac{D_T}{\tau}\right), \tag{6}$$

velocity of thermal wave propagation and

$$\lambda = v\tau, \tag{7}$$

where $\lambda$ is the mean free path of the heat carriers. With formula (6) equation (3) can be written as

$$\frac{1}{v^2}\frac{\partial^2 T}{\partial t^2} + \frac{1}{v^2\tau}\frac{\partial T}{\partial t} = \nabla^2 T. \tag{8}$$

From the mathematical point of view equation:

$$\frac{1}{v^2}\frac{\partial^2 T}{\partial t^2} + \frac{1}{D}\frac{\partial T}{\partial t} = \nabla^2 T$$

is the hyperbolic partial differential equation (PDE). On the other hand Fourier equation

$$\frac{1}{D}\frac{\partial T}{\partial t} = \nabla^2 T \tag{9}$$

and Schrödinger equation

$$i\hbar\frac{\partial \Psi}{\partial t} = -\frac{\hbar^2}{2m_i}\nabla^2 \Psi \tag{10}$$

are the parabolic equations. Formally with substitutions

$$t \leftrightarrow it, \ \Psi \leftrightarrow T. \tag{11}$$

Fourier equation (9) can be written as

$$i\hbar \frac{\partial \Psi}{\partial t} = -D\hbar \nabla^2 \Psi \tag{12}$$

and by comparison with Schrödinger equation one obtains

$$D_T \hbar = \frac{\hbar^2}{2m_i} \tag{13}$$

and

$$D_T = \frac{\hbar}{2m_i}. \tag{14}$$

Considering that $D_T = \upsilon^2 \tau$ (6) we obtain from (14)

$$\tau = \frac{\hbar}{2m_i \upsilon_h^2}. \tag{15}$$

Formula (15) describes the relaxation time for quantum thermal processes.

Starting with Schrödinger equation for particle with mass $m_i$ in potential $V$:

$$i\hbar \frac{\partial \Psi}{\partial t} = -\frac{\hbar^2}{2m_i} \nabla^2 \Psi + V\Psi \tag{16}$$

and performing the substitution (11) one obtains

$$\hbar \frac{\partial T}{\partial t} = \frac{\hbar^2}{2m_i} \nabla^2 T - VT \tag{17}$$

$$\frac{\partial T}{\partial t} = \frac{\hbar}{2m_i} \nabla^2 T - \frac{V}{\hbar} T. \tag{18}$$

Equation (18) is Fourier equation (parabolic PDE) for $\tau = 0$. For $\tau \neq 0$ we obtain

$$\tau \frac{\partial^2 T}{\partial t^2} + \frac{\partial T}{\partial t} + \frac{V}{\hbar} T = \frac{\hbar}{2m_i} \nabla^2 T, \tag{19}$$

$$\tau = \frac{\hbar}{2m_i \upsilon^2} \tag{20}$$

or

$$\frac{1}{\upsilon^2} \frac{\partial^2 T}{\partial t^2} + \frac{2m_i}{\hbar} \frac{\partial T}{\partial t} + \frac{2Vm_i}{\hbar^2} T = \nabla^2 T.$$

With the substitution (11) equation (19) can be written as

$$i\hbar \frac{\partial \Psi}{\partial t} = V\Psi - \frac{\hbar^2}{2m_i} \nabla^2 \Psi - \tau \hbar \frac{\partial^2 \Psi}{\partial t^2}. \tag{21}$$

The new term, relaxation term

$$\tau\hbar\frac{\partial^2\Psi}{\partial t^2} \qquad (22)$$

describes the interaction of the particle with mass $m_i$ with space-time. When the quantum particle is moving through the quantum void it is influenced by scaterring on the virtual electron=-positron pairs The relaxation time can be calculated as:

$$\tau^{-1} = \left(\tau_{e-p}^{-1} + ... + \tau_{Planck}^{-1}\right), \qquad (23)$$

where, for example $\tau_{e-p}$ denotes the scattering of the particle $m_i$ on the electron-positron pair ($\tau_{e-p} \sim 10^{-17}$ s) and the shortest relaxation time $\tau_{Planck}$ is the Planck time ($\tau_{Planck} \sim 10^{-43}$ s).

From equation (23) we conclude that $\tau \approx \tau_{Planck}$ and equation (21) can be written as

$$i\hbar\frac{\partial\Psi}{\partial t} = V\Psi - \frac{\hbar^2}{2m_i}\nabla^2\Psi - \tau_{Planck}\hbar\frac{\partial^2\Psi}{\partial t^2}, \qquad (24)$$

where

$$\tau_{Planck} = \frac{1}{2}\left(\frac{\hbar G}{c^5}\right)^{\frac{1}{2}} = \frac{\hbar}{2M_p c^2}. \qquad (25)$$

In formula (25) $M_p$ is the mass Planck. Considering Eq. (25), Eq. (24) can be written as

$$i\hbar\frac{\partial\Psi}{\partial t} = -\frac{\hbar^2}{2m_i}\nabla^2\Psi + V\Psi - \frac{\hbar^2}{2M_p}\nabla^2\Psi + \frac{\hbar^2}{2M_p}\nabla^2\Psi - \frac{\hbar^2}{2M_p c^2}\frac{\partial^2\Psi}{\partial t^2}. \qquad (26)$$

The last two terms in Eq. (26) can be defined as the *Bohmian* pilot wave

$$\frac{\hbar^2}{2M_p}\nabla^2\Psi - \frac{\hbar^2}{2M_p c^2}\frac{\partial^2\Psi}{\partial t^2} = 0, \qquad (27)$$

i.e.

$$\nabla^2\Psi - \frac{1}{c^2}\frac{\partial^2\Psi}{\partial t^2} = 0. \qquad (28)$$

It is interesting to observe that pilot wave $\Psi$ does not depend on the mass of the particle. With postulate (28) we obtain from equation (26)

$$i\hbar\frac{\partial\Psi}{\partial t} = -\frac{\hbar^2}{2m_i}\nabla^2\Psi + V\Psi - \frac{\hbar^2}{2M_p}\nabla^2\Psi \qquad (29)$$

and simultaneously

$$\frac{\hbar^2}{2M_p}\nabla^2\Psi - \frac{\hbar^2}{2M_p c^2}\frac{\partial^2\Psi}{\partial t^2} = 0. \tag{30}$$

In the operator form Eq. (21) can be written as

$$\hat{E} = \frac{\hat{p}^2}{2m_i} + \frac{1}{2M_p c^2}\hat{E}^2, \tag{31}$$

where $\hat{E}$ and $\hat{p}$ denote the operators for energy and momentum of the particle with mass $m_i$. Equation (31) is the new dispersion relation for quantum particle with mass $m_i$. From Eq. (21) one can concludes that Schrödinger quantum mechanics is valid for particles with mass $m_i \ll M_P$. But pilot wave exists independent of the mass of the particles.

For particles with mass $m_i \ll M_P$ Eq. (29) has the form

$$i\hbar\frac{\partial\Psi}{\partial t} = -\frac{\hbar^2}{2m_i}\nabla^2\Psi + V\Psi. \tag{32}$$

In the case when $m_i \approx M_p$ Eq. (29) can be written as

$$i\hbar\frac{\partial\Psi}{\partial t} = -\frac{\hbar^2}{2M_p}\nabla^2\Psi + V\Psi, \tag{33}$$

but considering Eq. (30) one obtains

$$i\hbar\frac{\partial\Psi}{\partial t} = -\frac{\hbar^2}{2M_p c^2}\frac{\partial^2\Psi}{\partial t^2} + V\Psi \tag{34}$$

or

$$\frac{\hbar^2}{2M_p c^2}\frac{\partial^2\Psi}{\partial t^2} + i\hbar\frac{\partial\Psi}{\partial t} - V\Psi = 0. \tag{35}$$

We look for the solution of Eq. (35) in the form

$$\Psi(x,t) = e^{i\,t}u(x). \tag{36}$$

After substitution formula (16) to Eq. (35) we obtain

$$\frac{\hbar^2}{2M_p c^2}\,^2 + \hbar + V(x) = 0 \tag{37}$$

with the solution

$$_1 = \frac{-M_p c^2 + M_p c^2\sqrt{1 - \dfrac{2V}{M_p c^2}}}{\hbar}$$

$$_2 = \frac{-M_p c^2 - M_p c^2\sqrt{1 - \dfrac{2V}{M_p c^2}}}{\hbar} \tag{38}$$

for $\dfrac{M_p c^2}{2} > V$ and

$$\begin{aligned}\omega_1 &= \dfrac{-M_p c^2 + i M_p c^2 \sqrt{\dfrac{2V}{M_p c^2} - 1}}{\hbar} \\ \omega_2 &= \dfrac{-M_p c^2 - i M_p c^2 \sqrt{\dfrac{2V}{M_p c^2} - 1}}{\hbar}\end{aligned} \qquad (39)$$

for $\dfrac{M_p c^2}{2} < V$.

Both formulae (38) and (39) describe the string oscillation, formula (27) damped oscillation and formula (28) over damped string oscillation.

### 3. Quantum – classical transition in the brain

As was shown in paragraph 2 the transition quantum – classical behavior occurs at mass $10^{-5}$ g, i.e. at the mass of the order of the human neuron mass – Planck mass.

The Planck mass depends on the Planck constant $\hbar$, light velocity $c$ and gravity constant $G$

$$M_P = \sqrt{\dfrac{\hbar c}{G}}. \qquad (40)$$

In the papers [13-15] the Klein – Gordon equation for Planck gas, i.e. gas of particles with mass was formulated and solved. As the one of the fundamental result – the time arrow creation by gravity was stated.

Considering the result of the present paper and the results of the papers [13-15] it can be argued that the gravitation plays the fundamental role in the brain operation:

1. The gravity, i.e. when $G \ne 0$, creates the arrow of time.
2. The gravity, through the Planck mass is responsible for the quantum – classical transition in the brain.

3. The brain can be considered as the fluid which consists of the Planck particles, i.e. the Planck fluid with very high velocity